\documentclass[useAMS,usenatbib]{mn2e}
\usepackage{graphics}
\usepackage{txfonts}

\newcommand{\gmas}{\mbox{$\dot{M_g}$}} %gas mass loss rate
\newcommand{\mlu}{\mbox{${\rm M}_{\odot}$\,yr$^{-1}$}}

\newcommand{\Msun}{\mbox{${\rm M}_{\odot}$}}\def\oversim#1#2{\lower0.5pt\vbox{\baselineskip0pt \lineskip-0.5pt
     \ialign{$\mathsurround0pt #1\hfil##\hfil$\crcr#2\crcr\sim\crcr}}}
\begin{document}
\title[Global gas and dust budget of the SMC]
{The global gas and dust budget of the Small Magellanic Cloud
%and comparisons with other gas and dust sources
}
%   \subtitle{Discrepancy of dust compositions between dust sources and ISM extinction curve}
\author[M. Matsuura, P.M.  Woods \& P.J. Owen]
{M.~Matsuura, Paul M.  Woods, P.J.  Owen
\\
  Department of Physics and Astronomy, University College London, 
	Gower Street, London WC1E 6BT, United Kingdom \\
             }

\date{Submitted}
%\date{Accepted. Received; in original form }
\pagerange{\pageref{firstpage}--\pageref{lastpage}} \pubyear{2012}

\maketitle
\label{firstpage}
\begin{abstract}
  In order to understand the evolution of the interstellar medium (ISM) of a galaxy,
  we have analysed the gas and dust budget of  the Small Magellanic Cloud (SMC). 
  Using the {\it Spitzer Space Telescope}, we measured the integrated gas mass-loss rate
  across asymptotic giant branch (AGB) stars and red supergiants (RSGs) in the SMC,
  and obtained a rate of 1.4$\times10^{-3}$\,\mlu. 
  This is much smaller than the estimated gas ejection rate from type II supernovae (SNe)
  (2--4$\times10^{-2}$\,\mlu).
  The SMC underwent a an increase in starformation rate  in the last 12\,Myrs,
  and consequently the galaxy has a relatively high SN rate at present.
  Thus, SNe are more important gas sources than AGB stars in the SMC.
  The total gas input from stellar sources into the ISM is 2--4$\times10^{-2}$\,\mlu. 
  This is slightly smaller than the ISM gas consumed by starformation
  ($\sim$8$\times10^{-2}$\,\mlu). 
  Starformation in the SMC relies on a gas reservoir in the ISM,
  but eventually the starformation rate will decline in this galaxy, unless gas infalls into the ISM
  from an external source.
  The dust injection rate from AGB and RSG candidates is 1$\times10^{-5}$\,\mlu.
  Dust injection from SNe is in the range of 0.2--11$\times10^{-4}$\,\mlu, although the SN contribution is rather uncertain.
  Stellar sources could be important for ISM dust ($3\times10^5$\,\Msun) in the SMC,
  if the dust lifetime is about 1.4\,Gyrs.
  We found that the presence of poly-aromatic hydrocarbons (PAHs) in the ISM cannot 
  be explained entirely by carbon-rich AGB stars.
  Carbon-rich AGB stars could inject only 7$\times10^{-9}$\,\mlu\, of PAHs at most,
  which could contribute up to 100\,\Msun\, of PAHs in the lifetime of a PAH.
  The estimated PAH mass of 1800\,\Msun\, in the SMC can not be explained.
  Additional PAH sources, or ISM reprocessing should be needed.
  
\end{abstract}

%
%________________________________________________________________
\large

%\tableofcontents

\section{Introduction}

Stars and the interstellar medium (ISM) of galaxies experience a constant exchange
of gas and dust.
Stars are formed in molecular clouds in the ISM, and evolve. Elements are synthesised
in stellar interior, and eventually they are ejected at the end of the stellar life. Gas ejected by stars
is enriched with metals, in comparison to with the gas where the stars were initially formed.
Dust grains are formed around evolved stars and supernovae (SNe) and they are injected into the ISM. 
They might be processed in the ISM, though that remains uncertain.
Eventually, dust grains are destroyed by shocks generated by SN blast winds.
The lifecycle of matter drives the evolution of the ISM, and ultimately, the evolution of galaxies.

The concept of the lifecycle of matter, in particular gas content, is well accepted 
and has been adopted in the chemical evolution models and stellar population models of galaxies
\citep[e.g.][]{Pagel:1998jm, Bruzual:2003ck, Kodama:1997vx}.
However,  it had been difficult to actually measure the gas and dust  feedback from stars into the ISM in real terms.
The {\it Spitzer Space Telescope} \citep{Werner:2004jt} has
provided an opportunity to measure
the gas feedback from stars in the Magellanic Clouds
\citep{Matsuura:2009fs, Srinivasan:2009bia, Boyer:2012vu},
delivering new constraints on chemical evolution models.

Dust is one of the important contents of galaxies.
Dust absorbs energy emitted from stars within a galaxy, and re-emits it in the infrared to sub-mm wavelengths.
The presence of dust controls the energy input and output from the ISM of galaxies, and affects 
the spectral shape of the galaxy and the underlining physics of the ISM \citep{Galliano:2005ei, Dunne:2011fr}.

Despite such an important role in the physics of the ISM and galaxies, it is still not well established
how dust mass in the ISM evolves
\citep{Sloan:2009ed, Draine:2009ur, Tielens:2005tb, Calura:2008eg}.
This requires detailed studies of the dust formation and destruction processes in stars and the ISM.
Dust grains are considered to be formed in a wide range of objects, particularly
stars in the late phase of evolution  \citep[e.g.][]{Gehrz:1989uy}.
Such theoretical studies have been conducted \citep{Nozawa:2003fd,
Morgan:2003cq, Ferrarotti:2006gj, Zhukovska:2008bw, Valiante:2009hg}
but are poorly constrained by observations at present.
Once grains are ejected from stars into the ISM, it has been proposed
that dust grains are re-processed and destroyed.
Dust grains might grow in the ISM, using stellar dust as seeds \citep[e.g.][]{Tielens:2005tb,Draine:2009ur},
but it is more difficult to directly measure these processes.
This paper focuses on establishing an understanding of what
the contribution of stellar dust is to the evolution of dust in the ISM.

One of the important dust sources are asymptotic giant branch (AGB) stars,
low- and intermediate-mass (1--8\,$\Msun$) evolved stars.
It has been well established that dust grains are formed in the AGB outflow  \citep[e.g.][]{Habing:1996wn}.
However, it is challenging
to make a quantitative analysis; namely to measure dust formed in evolved stars in entire populations across galaxies,
and evaluate their contribution to dust in the ISM, because  this requires a large survey of AGB stars
in a galaxy. One of the pioneering studies was based on the IRAS survey of the solar neighbourhood
\citep{Jura:1989br},
and the Spitzer Space Telescope opened up a possibility for such studies beyond the Milky Way \citep{Matsuura:2009fs, Srinivasan:2009bia}.

High-mass stars are considered to form dust in various evolutionary stages,
such as red supergiants (RSGs), Wolf-Rayet stars, luminous blue variables (LBVs), and supernova (SNe).
Due to limited number of sample, the mass dust formed in high mass stars, and in particular, SNe, is uncertain.
Recent {\it Spitzer} and {\it Herschel} studies of SNe and SN remnants (SNRs)
show that type-II SNe can form dust, but the reported dust masses range from $10^{-4}$ to $\sim$1\,\Msun\,
\citep{BenEKSugerman:2006gb, Meikle:2007gl, Barlow:2010cz, Matsuura:2011ij, Gomez:2012fm}.
Type-Ia SNe, which have an origin in low- and intermediate-mass binary stars,
appear to form very small amounts of dust  \citep{Gomez:2012jk}.

The explosion of high mass stars
creates expanding winds, which often have multiple velocity components  \citep{Kjaer:2010wl}.
When the fast blast wind collides with the ISM dust, ISM dust grains can be destroyed
via sputtering and shattering processes \citep{Jones:1996bi}.
The lifetime of dust is determined by its destruction by SN shocks.
Assuming  homogeneous gas and dust distributions, the lifetime of dust is estimated
to be nearly 1\,Gyr \citep{Jones:1996bi}.
Gas and dust are distributed inhomogenously in the ISM, so as  it  is very difficult
to estimate a precise dust lifetime.

In this paper, we measure the gas and dust injection rate from AGB stars and RSGs
in the Small Magellanic Cloud (SMC). This galaxy is only 56\,kpc away \citep{McCumber:2005jw}
and is a molecular-poor galaxy, yet
has some ongoing starformation.
The measured gas and dust masses ejected from AGB stars and RSGs
are compared with masses ejected from SNe, and  with the starformation rate in the SMC.
We discuss the current problems in understanding the gas and dust budget, and its evolution in these galaxies.

\section{Mass-loss rates of AGB stars and red supergiants} \label{mass-loss-rate}

Mass-loss rates of AGB stars and red supergiants correlate well with infrared excess.
Often near-infrared photometric points are used to represent the photospheric flux,
while mid-infrared fluxes illustrate the emission from the dust thermal emission.
Taking near- and mid-infrared colours is an indicator of mass-loss rates 
\citep{LeBertre:1998ty, Whitelock:1994tl}.

The correlation between colour and mass-loss rates have been established for
LMC and SMC carbon-rich AGB stars (\cite{Matsuura:2009fs},
who used the mass-loss measurements from \cite{Groenewegen:2007bi},
and photometric data from \citet{Meixner:2006eg} and \citet{Skrutskie:2006hl}).
The resultant relations were
\begin{equation} \label{eq-38}
\log\, \gmas =  -6.20 / ( ([3.6]-[8.0])+0.83) -3.39   
\end{equation}
in the range of $1<[3.6]-[8.0]<9$
\begin{equation} \label{eq-k8}
\log\, \gmas =  -14.50 / ( (K_s-[8.0])+3.86) -3.62  
\end{equation}
in the range of $1<K_s-[8.0]<9$, where $\gmas$ is the gas mass-loss rate.

The relation for $K_s-[24]$ for the same data set is found to be
\begin{equation}\label{eq-k24}
 log\, \gmas =  -20.25 / ( (K_s-[24])+4.98) -3.47,
\end{equation}

\begin{figure}
\centering
\resizebox{0.8\hsize}{!}{\includegraphics*[27,296][429,731]{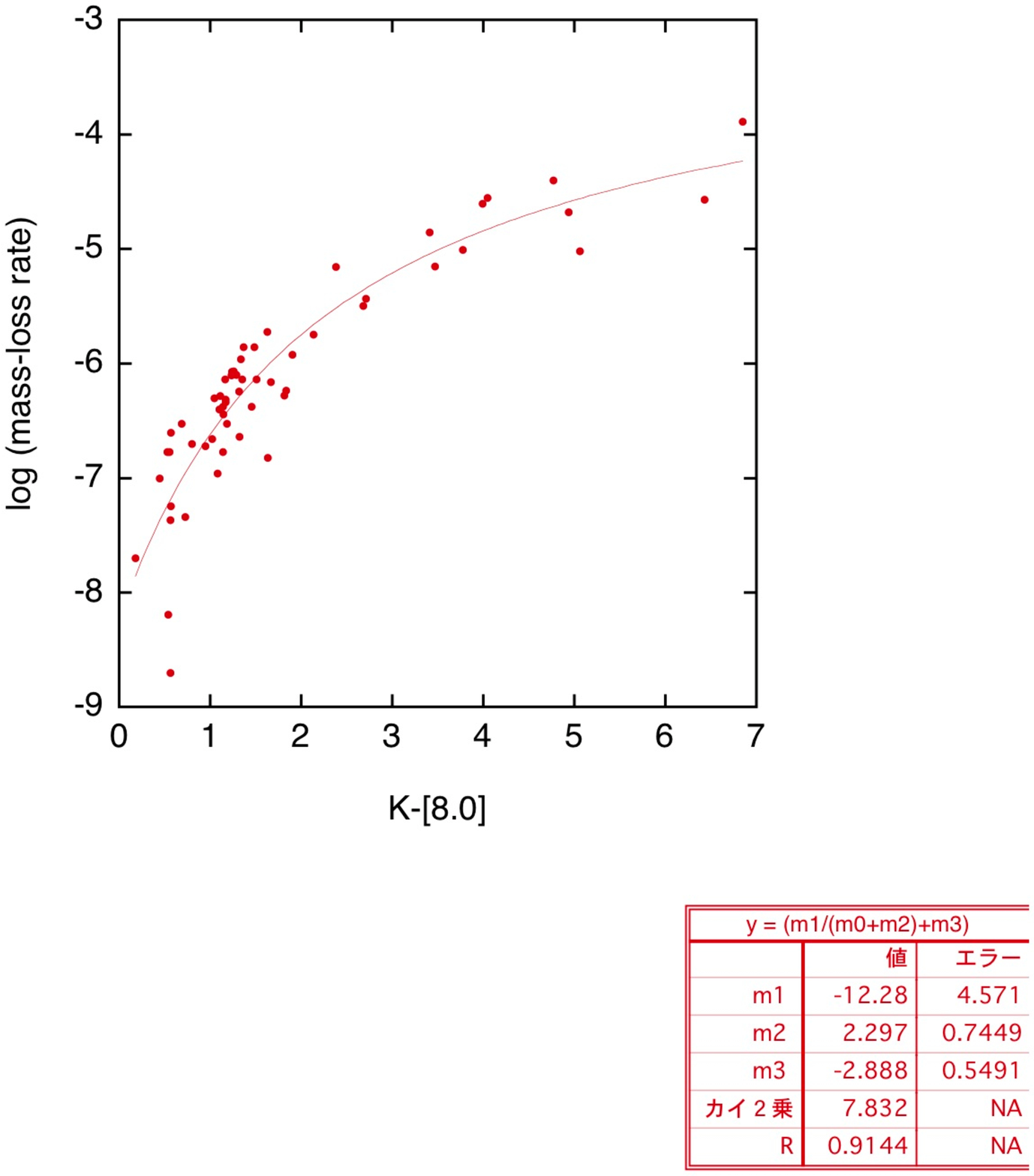}}
\resizebox{0.85\hsize}{!}{\includegraphics*[27,296][429,731]{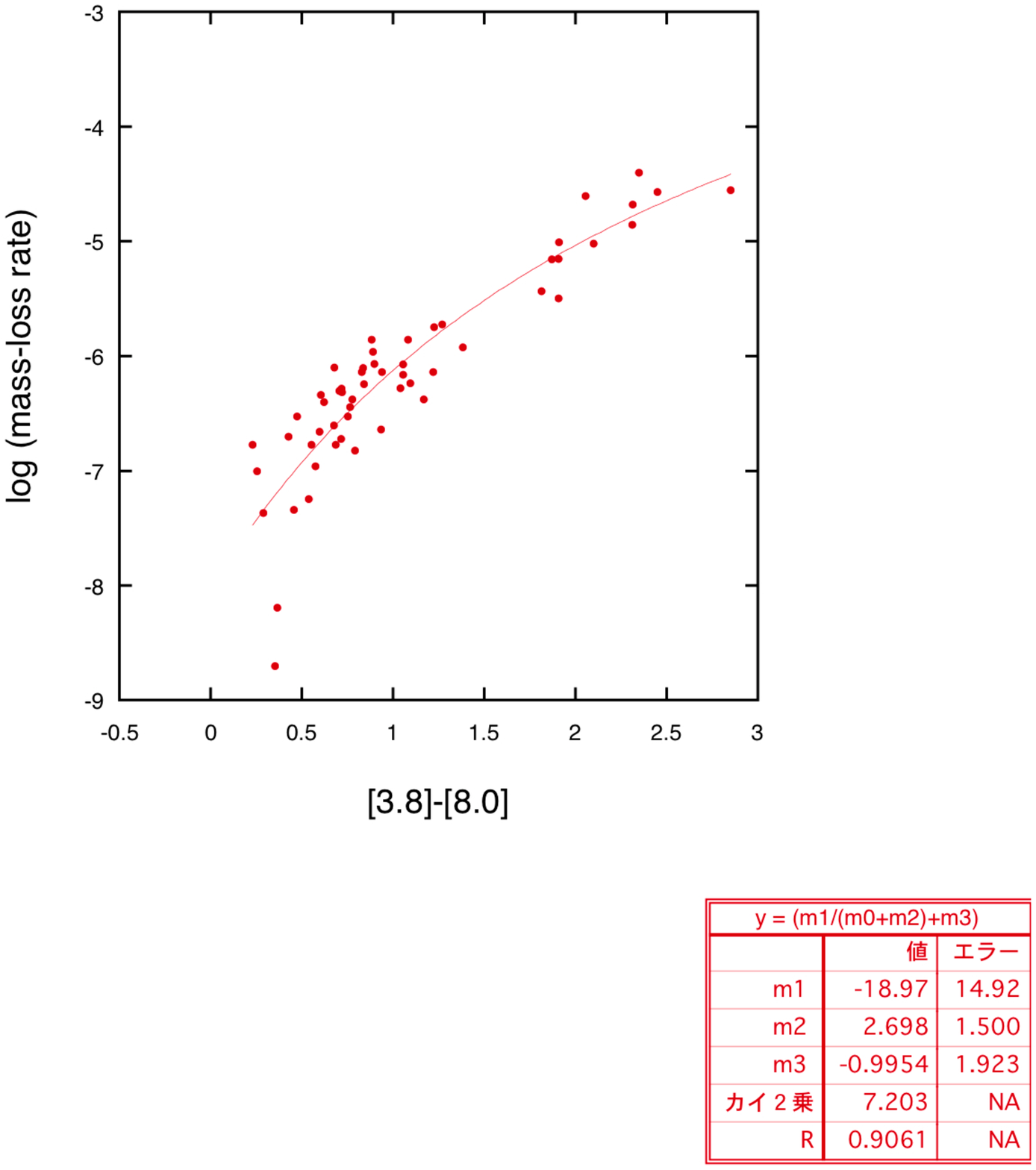}}
\resizebox{0.85\hsize}{!}{\includegraphics*[27,296][429,731]{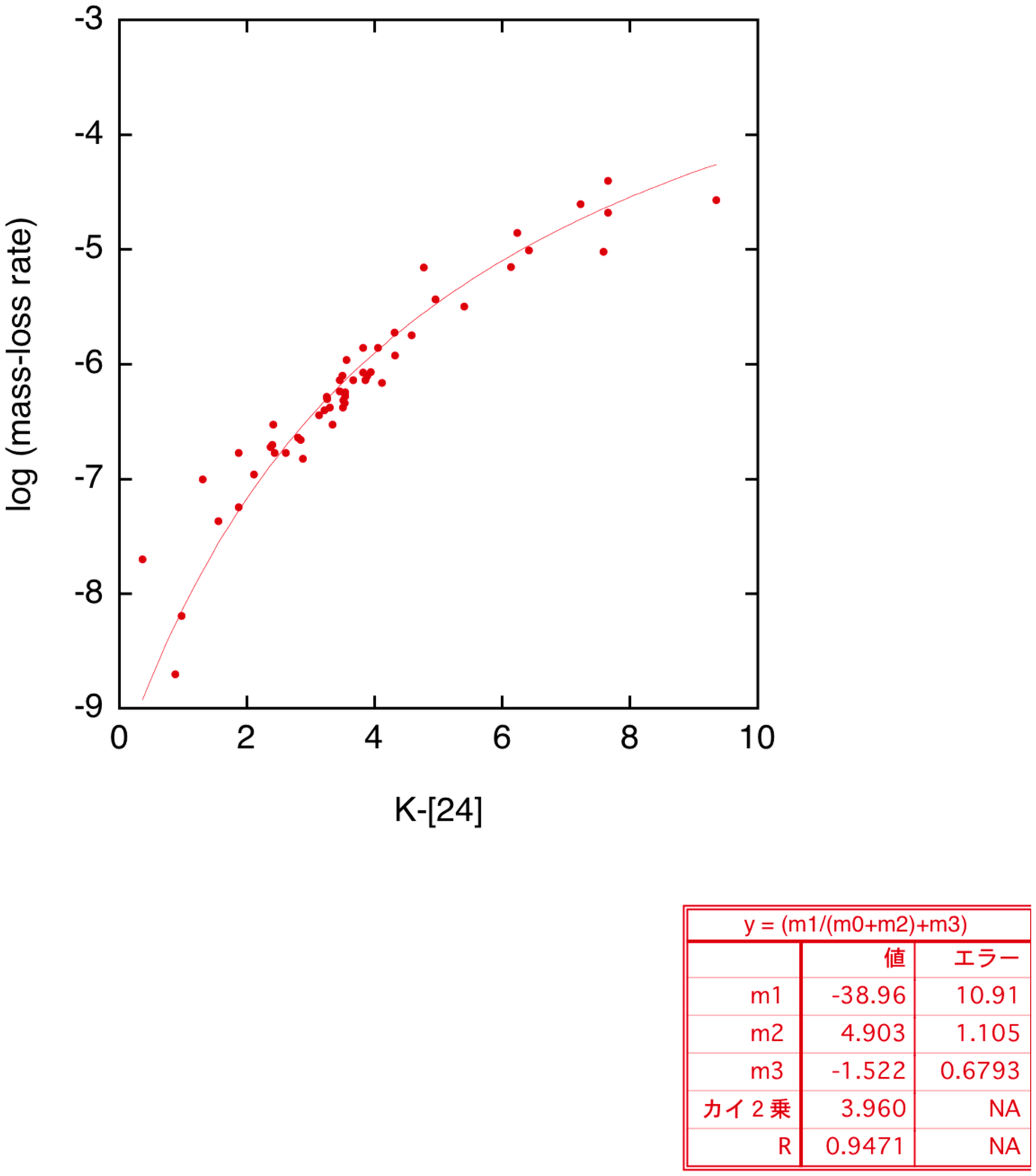}}
\caption{Correlation between mass-loss rate (in $\mlu$) and colour for oxygen-rich AGB stars and RSGs
\label{Fig-mass-loss}}
\end{figure}

We further derive the relationship for oxygen-rich stars, using
\citet{Groenewegen:2009jq}'s analysis of
oxygen-rich AGB stars and red supergiants
in the Magellanic Clouds. We correlate their photometric data with Spitzer and 2MASS data
\citep{Gordon:2011jq}.
The resultant correlation is found in Fig.\ref{Fig-mass-loss},
and the fits to these data are given as 
\begin{equation} \label{eq-38-o}
\log\, \gmas =  -18.97 / ( ([3.6]-[8.0])+2.698) -0.9954   
\end{equation}
in the range of $1<[3.6]-[8.0]<3$
\begin{equation} \label{eq-k8-o}
\log\, \gmas =  -12.28 / ( (K_s-[8.0])+2.297) - 2.888 
\end{equation}
in the range of $1<K_s-[8.0]<7$, and
\begin{equation}\label{eq-k24-o}
 log\, \gmas =  -38.96 / ( (K_s-[24])+4.903) -1.522,
\end{equation}
in the range of $1<K_s-[24]<9$.

It is difficult to measure the mass-loss rates of stars with very little infrared excess 
and blue colour cut-offs.
However, contributions from very low-mass loss rate stars are not important for the overall global gas and dust budget
\citep{LeBertre:2001ho, Matsuura:2009fs}.

%_________________________________________________________________
%_________________________________________________________________

\section{Object classifications}

The first step is to understand the infrared  photometric survey and classify the point sources.
Colour magnitude diagrams (CMDs) and colour-colour diagrams (CCDs) are commonly used
to classify the point sources \citep[e.g.][]{Blum:2006ib, Ita:2008vl, Woods:2010it}.
In this work, we use the LMC objects to set object classifications based on a CMD and a CCD, 
and apply the classification method to SMC objects.

In our previous study, we focused on classification based on
[3.6]$-$[8.0] v.s. [8.0] CMD to extract carbon-rich AGB stars
\citep{Matsuura:2009fs}. We are going to revise this classification,
and particularly optimise the classifications for high mass-loss rate AGB stars,
which are dominant for the gas and the dust budget.

\subsection{Cross-identifications}

We start the object classification process by cross-identifying 
the LMC photometric data with spectroscopically known objects in the LMC.

\citet{Matsuura:2009fs} have assembled the LMC objects with spectroscopical classifications
and we use these sample:
Sample of M-type giants and supergiants, carbon-rich stars, planetary nebulae are assembled from
\citet{Kontizas:2001ck},  \citet{Cioni:2001gq},  \citet{Sanduleak:1977wx},  \citet{Blanco:1980bu},
\citet{Westerlund:1981wi},  \citet{Wood:1983bh},  \citet{Wood:1985jm},
\citet{Hughes:1989cr}
and  \citet{Reid:1988vj}.
We added samples of Wolf Rayet (WR) stars  \citep{Breysacher:1999tx}
and 
S\,Dor variables 
 \citep[Luminous Blue Variables (LBVs) ; ][]{VanGenderen:2001kk},
 post-AGB stars 
\citep{Gielen:2009iw, Volk:2011iw, Matsuura:2011us}, and 
planetary nebulae \citep{Woods:2010it,  Matsuura:2011us}.

\subsection{Object classifications}

\subsubsection{LMC colour-magnitude diagram}

Figure\,\ref{Fig-38-8} shows the [3.8]-[8.0] vs [8.0] CMDs of the LMC objects.
This diagram was used for object
classifications of carbon-rich AGB stars and oxygen-rich AGB stars/red supergiants
in our previous study \citep{Matsuura:2009fs}.
The black lines shows the separation of these two types.
LMC Spitzer spectroscopic observations \citep{Kemper:2010bw}
confirm that this classification is effective \citep{Woods:2010it}.

This CMD classification is simple and largely correct.
Actually, contamination of a small number of red objects (carbon-rich post-AGB stars and PNe and distant galaxies)
into AGB stars and RSGs could potentially change the analysis of the global gas and dust budget, which is the final
aim of this paper.
The further and more severe problem is contamination of a few, but high mass-loss rate oxygen-rich AGB stars
which fall into the carbon-rich AGB region at about [3.6]$-$[8.0]$\sim$2 and [8.0]$\sim$6\,mag.
We introduce one more step in the analysis, as described in Sect.\,\ref{sect-CCD},
to minimise these contaminations.

Many point sources are found in the region between  $2<[3.6]-[8.0]<4$  and $[8.0]<12$.
This is associated with distant galaxies \citep{Koziowski:2009hp}.

\subsubsection{LMC colour-colour diagram}\label{sect-CCD}

%_________________________________________________________________
\begin{figure*}
\centering
\rotatebox{270}{
\begin{minipage} {11cm}
\resizebox{\hsize}{!}{\includegraphics*[93,156][518,718]{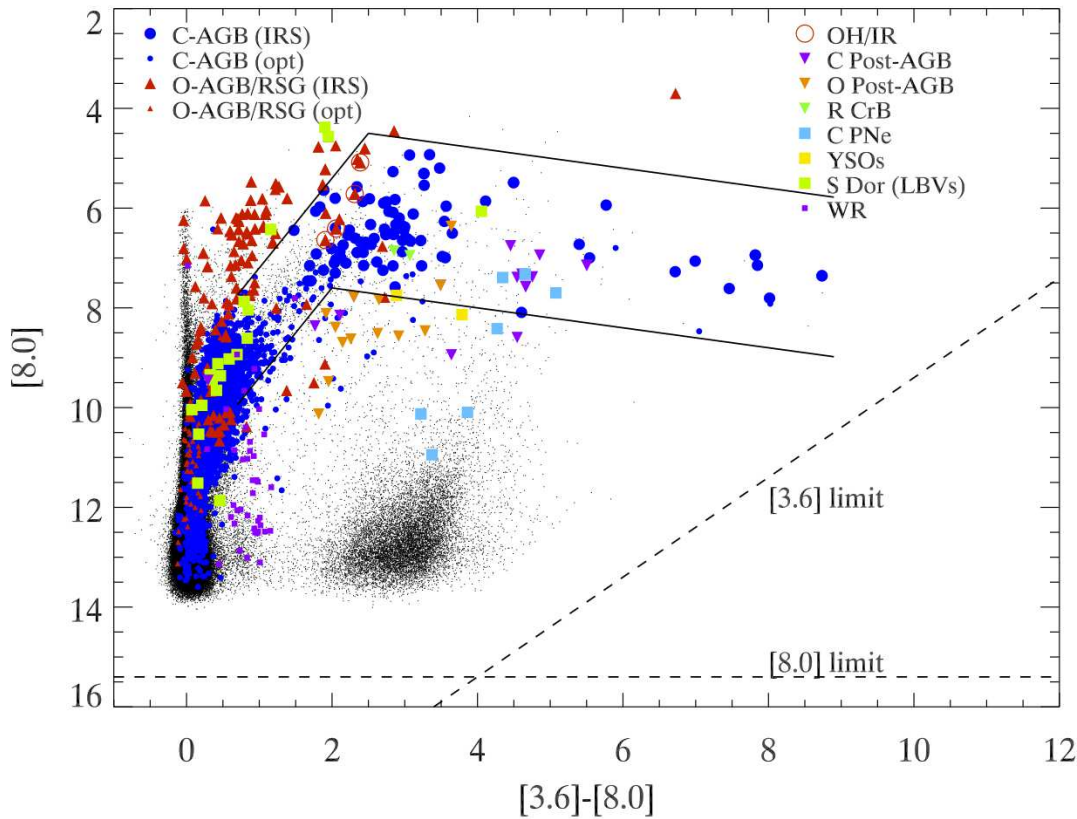}}
\end{minipage}}
\caption{The infrared colour-magnitude diagram of the LMC objects.
 Dots represent all point sources obtained from the SAGE catalogue \citep{Meixner:2006eg}.
 Circles show the spectroscopically identified carbon-rich AGB stars.
 Upword-pointing triangles show oxygen-rich AGB stars and red supergiants (RSGs).
 In addition to the AGB stars, S\,Dor type variables (luminous blue variables),
 and Wolf Rayet (WR) stars are plotted.
 The dashed lines show the expected detection limit of the final SAGE catalogue.  
\label{Fig-38-8}}
\end{figure*}
%_________________________________________________________________

\begin{figure*}
\centering
\rotatebox{270}{
%\begin{minipage} {6.5cm}
\begin{minipage} {11cm}
\resizebox{\hsize}{!}{\includegraphics*[93,156][516,718]{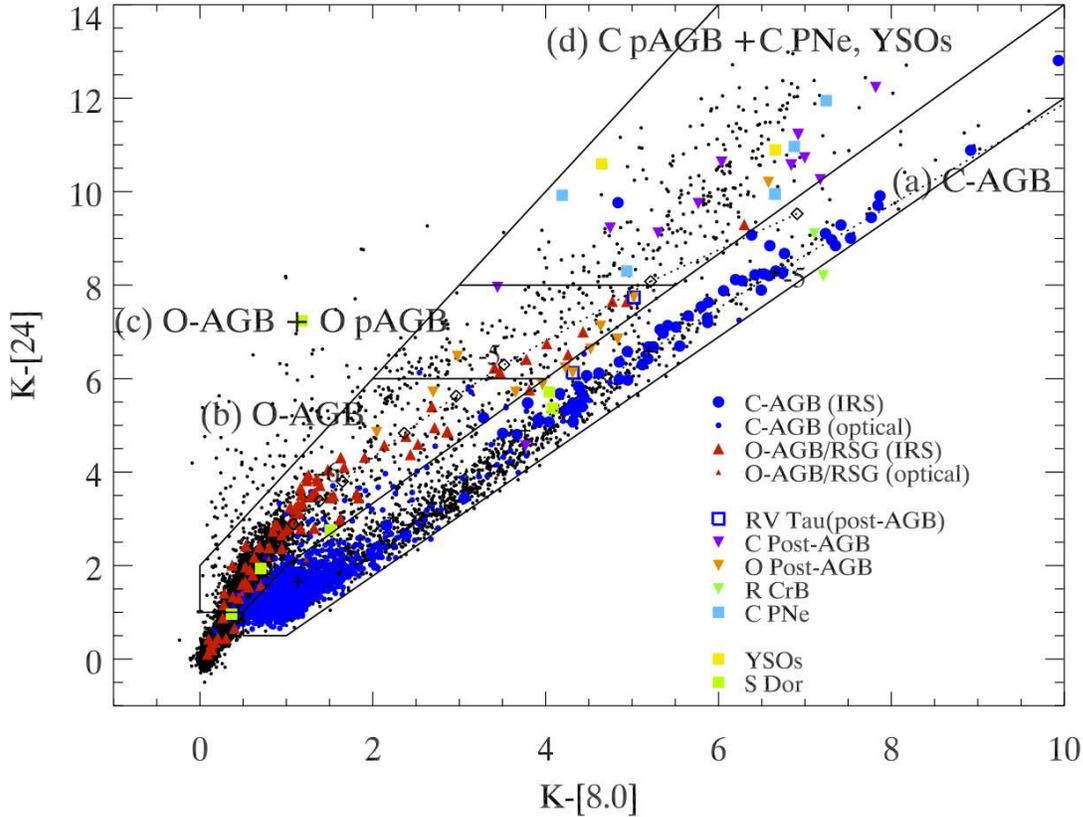}}
\end{minipage}}
\caption{Colour-colour diagram of the point sources in the LMC.
 The objects with spectroscopic  classifications are marked with symbols.
 We divide the colour-colour diagram into 4 regions (a)--(d) and dominant types of objects 
 in these regions are indicated.
 Oxygen-rich stars and carbon-rich stars follow two separat sequences.
 Two dashed lines show the estimated mass-loss rate and colour relations for oxygen-rich
 and carbon-rich stars, respectively, and they are calculated from equations (1)--(6). 
 The numbers alongside the lines show the mass-loss rates
 in log-scale (Sect.\ref{mass-loss-rate}).
\label{Fig-K8-K24}}
\end{figure*}

In order to improve object classifications further, we introduce another colour-colour diagram, 
the $K-[8.0]$ vs $K-[24]$ of the LMC objects, as plotted in Fig.\,\ref{Fig-K8-K24} .

Prior to the plot, point sources associated with distant galaxies \citep{Koziowski:2009hp} were removed. 
We defined their distribution as $[3.6]-[8.0]>2.0$ and $[8.0]<10$ from Fig.\,\ref{Fig-38-8}. 
These galaxies would appear approximately between $1.5< K-[8.0] <2.5$ and $5<K-[24]<7$ in the CCD,
if they had been plotted.

 We overlay the mass-loss vs colour relationship derived in Sect.\,\ref{mass-loss-rate} 
on top of the CMD. There are two separate sequences for oxygen-rich and carbon-rich objects.
The redder the colour, the higher the mass-loss rate, in general, as the equations show.

We use this CCD and the CMD to classify objects. We cross check the reliability of these
classifications with SIMBAD.
We found that the contaminations are not a big issue for the LMC gas and dust
budget, but it does change slightly (10\%) for the SMC.
A cross-check with SIMBAD found that non-AGB red objects that contaminated the AGB stars are, 
if they have known identifications, PNe \citep{Henize:1955gp} and YSO candidates \citep{Bolatto:2007hh}.

%_________________________________________________________________
%_________________________________________________________________

\subsubsection{SMC colour-magnitude diagram}

Once we establish the object classifications using the LMC objects,
we apply the same method to the SMC objects.

\begin{figure*}
\centering
\rotatebox{270}{
%\begin{minipage} {6.5cm}
\begin{minipage} {11cm}
\resizebox{\hsize}{!}{\includegraphics*[93,156][516,718]{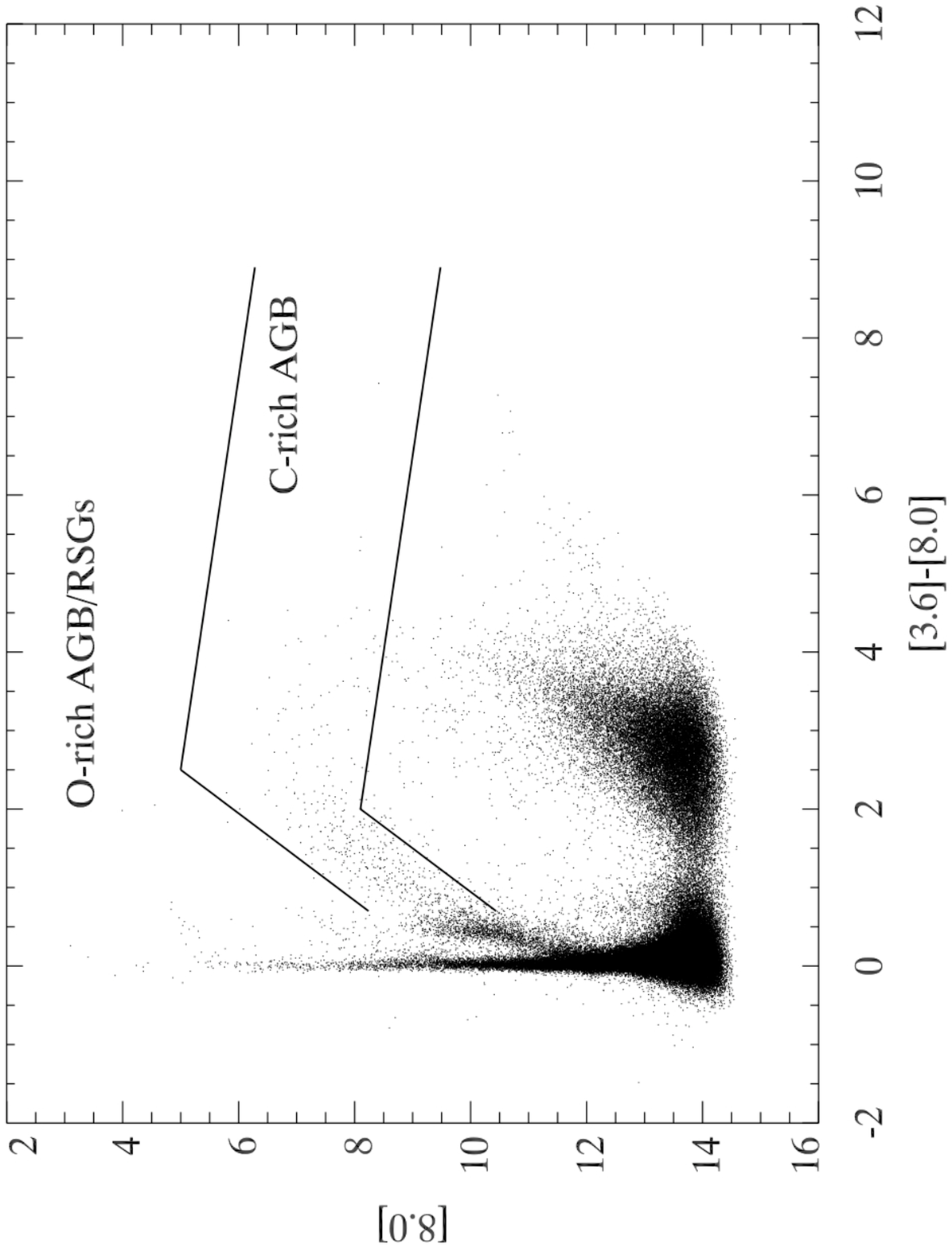}}
\end{minipage}}
\caption{The colour-magnitude diagram of the SMC objects.
The regions which are expected to be dominated by carbon-rich AGB stars and oxygen-rich AGB stars and red supergiants are indicated. 
\label{Fig-38-8-smc}}
\end{figure*}

Figure\,\ref{Fig-38-8-smc} shows the CMD of SMC point-sources extracted from
the SAGE-SMC catalogue \citep{Gordon:2011jq}.
The black lines show the separation between carbon-rich AGB stars and oxygen-rich AGB stars/red supergiants,
which was derived from LMC objects, but 
are scaled by the difference of distance moduli between these two galaxies.
The distance moduli of the LMC and the SMC are 18.5 and 18.9\,mag respectively
\citep{Nikolaev:2000ex, Westerlund:1990uh},
so that the separation line was shifted by 0.4 magnitude fainter in [8.0] for the SMC objects.

There is a clear difference in LMC and SMC CMDs:
fewer red stars are found in the SMC than the LMC.
Similarly, there are only three carbon-rich candidates with very red colours
([3.6]$-$[8.0]$>$5.0), whereas there are 33 such stars in the LMC.
We take into account the difference in the number of  stars in these two galaxies.
Taking the absolute $V$-band magnitude as a measure of number of stars in a galaxy,
the SMC should have about 7 times fewer stars than the LMC
\citep{Westerlund:1990uh}.
The difference in the red carbon-rich AGB stars appears be slightly larger than the difference
in the number in the stars, but this is not conclusive.

%_________________________________________________________________
%_________________________________________________________________

\subsubsection{SMC colour-colour diagram}

\begin{figure*}
\centering
\rotatebox{270}{
%\begin{minipage} {6.5cm}
\begin{minipage} {11cm}
\resizebox{\hsize}{!}{\includegraphics*[93,156][516,718]{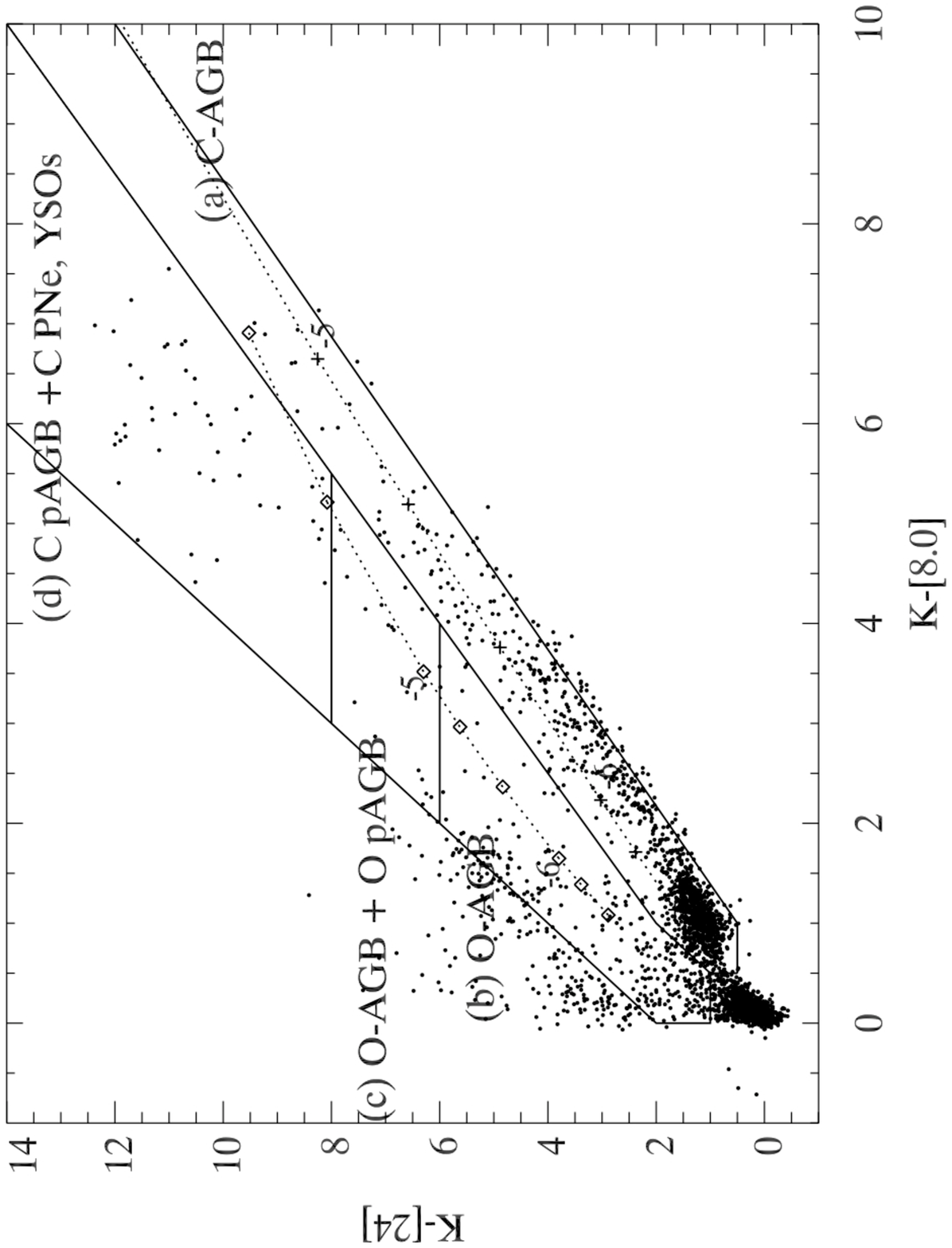}}
\end{minipage}}
\caption{Same as Fig.\,\ref{Fig-K8-K24}, but for the point sources in the SMC.
 There are very few oxygen-rich stars with colour redder than $K-[8.0]>1.8$, 
 which corresponds to a mass-loss rate
 higher than $>10^{-6}$\,\Msun\,yr\,s$^{-1}$. 
\label{Fig-K8-K24-smc}}
\end{figure*}

Figure\,\ref{Fig-K8-K24-smc} shows the same combination of 
CCD as Fig.\,\ref{Fig-K8-K24}, but for the Small Magellanic Cloud (SMC).
There is a slight difference in the distributions of stars compared with the LMC diagram:
the SMC has very few oxygen-rich AGB stars with colour redder than $K-[8.0]>1.8$.
This corresponds to approximately a mass-loss rate higher than $10^{-6}$\,\Msun\,yr$^{-1}$.

The SMC CCD shows that 
there are a reasonable number of stars found in region (a), which are carbon-rich AGB candidates.
Oxygen-rich AGB stars, represented by region (b), are relatively  scarce.
There are even fewer oxygen-rich AGB stars in the SMC than the LMC (Fig.\ref{Fig-K8-K24}).

%_________________________________________________________________
%_________________________________________________________________

\section{Gas and dust injection rate from evolved stars}

%_________________________________________________________________
\begin{table*}
% \centering
  \caption{ Gas and dust injected into the ISM of the LMC and the SMC \label{Table-budget}}
\begin{center}
 \begin{tabular}{lrrrrrrrrccccccccc}
  \hline
 & \multicolumn{2}{c}{LMC} & \multicolumn{2}{c}{SMC} \\
Sources & \multicolumn{1}{c}{Gas} & \multicolumn{1}{c}{Dust}  & \multicolumn{1}{c}{Gas} & \multicolumn{1}{c}{Dust} \\
 & ($10^{-2}$\,\mlu)& ($10^{-5}$\,\mlu)&  ($10^{-2}$\,\mlu)&  ($10^{-5}$\,\mlu)& \\ \hline
Carbon-rich AGB stars  & 0.7  & 4   & 0.08 & 0.4  \\
Oxygen-rich AGB + RSGs & 0.8  & 4   & 0.06 & 0.3  \\
Type II SNe            & 6--13 & 7--400 & 2--4 & 2-110 \\
WR stars              & $\sim0.1 $&  & $\sim 0.01$ & \\
OB stars               & 0.1--1?  & &  $\sim$0.03-0.3 & \\ \hline
Star-formation rate    & 20--30 &  & 8 \\
\hline
\end{tabular}
\label{Table-injection}
\end{center}
\end{table*}
%________________________________________________________________

\subsection{Integrated mass-loss rate from AGB stars and red supergiants}

Combining the information from object classifications and mass-loss rates from the previous sections,
we can estimate the integrated gas and dust mass-loss rates from AGB stars and red supergiants.
This is the total gas and dust mass injected from evolved stars into the ISM.

In order to estimate mass-loss rates of individual AGB stars and RSGs,
we use the correlation of  [3.6]$-$[8.0] with mass-loss rate. 
An alternative choice is $K_s-$[8.0], but some high mass-loss rate
stars are not detected at 2MASS $K_s$-band, and  these stars could contribute a significant fraction 
to the integrated mass-loss rate from evolved stars.

Object classification is based on [8.0] vs. [3.6]$-$[8.0] colour, and we supplementary use
$K_s-$[8.0] vs $K_s-$[24] to remove the contamination of distant galaxies in the red carbon-rich AGB region,
as well as high mass-loss rate oxygen-rich AGB stars and RSGs in the carbon-rich AGB region.

By integrating the mass-loss rates of individual AGB stars, we can estimate the total gas and dust inputs from 
AGB stars and red supergiants into the ISM 
of the SMC, as summarised in Table\,\ref{Table-budget}.
The contribution from oxygen-rich AGB stars and RSGs contribution is 0.06$\times10^{-2}$\,\Msun\,yr$^{-1}$,
while carbon-rich AGB stars contribute 
0.08$\times10^{-2}$\,\Msun\,yr$^{-1}$ in gas.
The total gas injection rate from AGB stars and red supergiants is 0.1$\times10^{-2}$\,\Msun\,yr$^{-1}$ for the SMC.

The estimated total gas input from evolved stars is 2$\times10^{-2}$\,\Msun\,yr$^{-1}$ for LMC.
This includes both oxygen-rich and carbon-rich AGB stars and RSGs. 
In our previous work on the LMC \citep{Matsuura:2009fs},  
the integrated mass-loss rate from oxygen-rich AGB stars and RSGS
was a crude estimate based on previously known surveys before the SAGE.
Tthe current estimation has a slightly higher value.
The integrated mass-loss rate is added up, using the contribution from  
carbon-rich AGB stars  (0.7$\times10^{-2}$\,\Msun\,yr$^{-1}$)
and oxygen-rich AGB stars and RSGs (0.8$\times10^{-2}$\,\Msun\,yr$^{-1}$).

In the SMC, dust inputs, which are directly measured from mid-infrared observations,
are 0.4$\times10^{-5}$\,\Msun\,yr$^{-1}$ for carbon-rich AGB
and 0.3$\times10^{-5}$\,\Msun\,yr$^{-1}$ for oxygen-rich AGB and RSGs.

The gas-to-dust mass ratio is assumed to be 200 for all AGB stars and RSGs 
as in \citet{Groenewegen:2007bi, Groenewegen:2009jq}.
This is the largest uncertainty in our estimate of the integrated gas mass-loss rate.
The ratio of 200 is a widely assumed value for the Galactic AGB stars and RSGs,
but it is not clear that this ratio is applicable to extra-Galactic evolved stars,
where the metallicities of galaxies are different.
The gas-to-dust ratio in the ISM and evolved stars is left for future investigations, 
such as  ALMA observations.

\citet{Boyer:2012vu} has estimated that the total dust injection rate is $1\times10^{-7}$\,\mlu\,
from carbon-rich AGB stars, $6\times10^{-7}$\,\mlu\, from ``extreme-'' AGB stars,
which they expected to be mostly carbon-rich, and 0.7$\times10^{-8}$\,\mlu\, from oxygen-rich AGB stars,
and 3$\times10^{-8}$\,\mlu\, from RSGs.
We removed the ambiguity of ``extreme'' from \citet{Boyer:2012vu}'s classification.
We found that the contribution from oxygen-rich AGB stars and red supergiants are much larger than their estimate.
\citet{Wood:1992ih} found three OH/IR stars in the SMC.
They have not estimated the mass-loss rate of these particular three objects,
but the typical range of mass-loss rate of OH/IR stars are $3\times10^{-5}$--$3\times10^{-4}$\,\mlu \citep{Wood:1992ih} .
\citet{Wood:1992ih} estimated these mass-loss rates based on infrared observations,
and they adopted a gas-to-dust ratio of 200. The sum of the dust mass-loss rates from these three OH/IR stars alone
is estimated to be $2\times10^{-7}$--$2\times10^{-6}$\,\mlu.
\citet{Groenewegen:2009jq} analysed the mass-loss rates of AGB stars and RSGs in Magellanic Clouds, 
and 11 MSX named oxygen-rich SMC stars
have a total dust mass loss rate of $4\times10^{-7}$\,\mlu.
These values show that our estimates are reasonable.
This does not affect \citet{Boyer:2012vu}'s overall conclusion that AGB stars and RSGs are not the dominant source of
dust in the ISM, which we also found.

\subsection{SN gas and dust injection rate}

In order to estimate the average gas and dust injection rate from core-collapse SNe into the ISM,
we need to estimate  SN rates and the progenitor mass, as well as dust mass formed in the SNe themselves.

The SN rates have been estimated for both the LMC and the SMC,
using the number and age of the supernova remnants (SNRs) in these galaxies as a starting point.
In our previous LMC study \citep{Matsuura:2009fs}, we used 
SN rates from \citet{Mathewson:1983gk} and \citet{Filipovic:1998vm}.
The \citet{Mathewson:1983gk} estimate is a factor of two lower than \citet{Filipovic:1998vm}.
\citet{Filipovic:1998vm}'s study is more recent, so we adopt this value here. 
Their rate includes both type Ia and II SNe.
\citet{Tsujimoto:1995wl} estimated the ratio of type Ia over type II is 0.2--0.3 for both the LMC and the SMC.
Combining the SN rate from \citet{Filipovic:1998vm} and type Ia and type II ratio from \citet{Tsujimoto:1995wl},
we obtain type SN II rate of one in every 125--143\,years in the LMC and one in every  438--500\,years in the SMC.

In our previous study of the LMC \citep{Matsuura:2009fs},
we have used the average SN progenitor mass of 8\,\Msun\, from 
SN progenitor surveys in the galaxies \citep{Smartt:2009ge}.
Eight solar masses correspond to the lowest range of mass that ends their stellar evolution as SNe.
One possibility is that this average progenitor mass might have some bias towards the lowest end of SN progenitors,
because they are more numerous than higher mass stars and more frequently detected.
We take this possibility into consideration.

A mean mass of high-mass stars is much higher than 8\,\Msun;
if we take a mean mass of the initial mass
function \citep{Kroupa:2001vq} with the high mass cut off of 50\,\Msun, the  average mass of high-mass stars
would be about 16\,\Msun. 
We take this to give a maximum possible mass of gas ejection rate per SN.

Considering the progenitor mass range and SN rate,  
the integrated gas inputs from type II SNe are approximately,
6--13$\times10^{-2}$\,\Msun\,yr$^{-1}$ for the LMC, 
and 2--4$\times10^{-2}$\,\Msun\,yr$^{-1}$ for the SMC.

SNe form dust in their ejecta, however, measured dust masses from SNe and SNRs have a wide range.
Mid-infrared observations found the lower limit of dust mass to be about 
$\sim10^{-5}$\,\Msun\, \citep{Meikle:2007gl}, but
up to $\sim10^{-2}$\,\Msun\,\citep{BenEKSugerman:2006gb, Fox:2011kg}.
Recently, the Herschel Magellanic Survey 
\citep[HERITAGE; ][]{Meixner:2010kj} found 
a large amount of dust in SN\,1987A  \citep{Matsuura:2011ij},
and its mass is about 0.3--0.7\,\Msun, showing SNe can form significant dust mass,
following a measurements in Galactic SNR, Cas\,A \citep[0.08\,\Msun\, of dust; ][]{Barlow:2010cz, Sibthorpe:2010bb}.
It seems that the measured dust mass does not correlate with the progenitor mass
\citep{Otsuka:2011gc, Gall:2011hr}.
In our estimate of the SN dust injection rate, we take two cases, $\sim10^{-2}$\,\Msun\, as a conservative case,
as taken in our previous study \citep{Matsuura:2009fs}.
The other case is 0.5\,\Msun\, per SNe.
This assumes that the dust detected in SN\,1987A after 25 years from the explosion
would not be destroyed in the later stage by the ISM and SN wind collisions.
The dust mass per event brings
the dust ejection rate to
7--400$\times10^{-5}$\,\Msun\,yr$^{-1}$ for the LMC, 
and 2--110$\times10^{-5}$\,\Msun\,yr$^{-1}$ for the SMC.

\subsection{Other dust and gas sources}

There are other possible sources which could contribute to the gas and dust budgets,
such as Wolf-Rayet (WR) stars, luminous blue variables (LBVs), novae and OB stars.
We briefly discuss their contributions to the budget.

Studies of Galactic WR stars show that
late-type (WR9 and WC 10) stars  form carbonaceous dust \citep{Williams:1987wj}.
In the Magellanic Clouds, such late-type WR stars have not  been found yet.
Among known WR stars,  the LMC WR star (HD 36402;  WC 4) was detected at mid-infrared wavelength
\citep{Williams:2011wr}, and its mass is estimated to be $1.5\times10^{-7}$\,\Msun.
It is not clear what the time scale of this dust formation and ejection rate is, but WR dust input
is much smaller than that of AGB stars and SNe combined in the LMC.
There is no report of dust formation in SMC WR, so far.

There are just over ten WR stars known in the SMC \citep{Pasemann:2011te}.
Their mass loss is line driven, and the mass-loss rate decreases with metallicity \citep{Kudritzki:2002jd, Crowther:2006go}.
Assuming each WR star ejects approximately $1\times10^{-5}$\,\mlu\, of gas \citep{Crowther:2000wr},
WR stars eject is about $1\times10^{-4}$\,\mlu\, of gas into the ISM in the SMC.
The LMC WR catalog lists 134 stars \citep{Breysacher:1999tx} and about 
$1\times10^{-3}$\,\mlu\, of gas is ejected from WR stars into the ISM in the LMC.

There are over 300 OB stars in the SMC \citep{Lamb:2011te}.
If their mass-loss rate is typically about $10^{-8}$--$10^{-6}$\,\mlu\, per star \citep{Prinja:1987uv},
the total mass-loss rate is about $\sim3\times10^{-5}$--$\sim3\times10^{-3}$\,\mlu.
In the LMC, the total mass-loss rate from approximately 1000 OB stars \citep{Bastian:2009ia} is
about $\sim1\times10^{-4}$--$\sim1\times10^{-2}$\,\mlu.

Luminous blue variables 
(LBVs) could potentially make significant amounts of silicate dust \citep{Morris:1999du, Gomez:2010gz}.
Among the S\,Dor star catalog of \citet{VanGenderen:2001kk},
three LBVs  have an infrared excess in the LMC,
which are R71, R127 and BAT 99-83 \citep{Bonanos:2009fk},
 indicating  dust formation in these stars.
In the SMC \citet{Bonanos:2010jk} found three LBVs that indicate an infrared excess.
However, we found that solely IR data can not distinguish between RSGs and LBVs (Fig.\ref{Fig-38-8}),
so that LBVs are combined together with RSGs in the overall gas and dust budget.

Classical novae can form dust and observations of Galactic novae reported dust mass
ranges between $10^{-10}$--$10^{-5}$\,\Msun\, depending on the objects \citep{Gehrz:1998jk}. As far as we are aware,
there is no report of dust formation in novae in the Magellanic Clouds, so that we do not include novae 
in the dust budget, at this moment.

\citet{Kastner:2006hs} found two LMC Be stars to have dust in discs.
These are probably formed during occasional outbursts.
The estimated mass-loss rate was $5\times10^{-4}$\,\mlu\,
giving the minimum mass injection rate from Be stars into the ISM of $10^{-3}$\,\mlu.
\citet{Bonanos:2010jk} have cross correlated optical spectral surveys of massive stars in the SMC  \citep{Evans:2004hm},
and found that five B[e] stars have infrared excess. They could have contributed approximately
$1\times10^{-5}$\,\mlu\, of dust.

Two R\,CrB stars have an infrared excess.
One is MSX LMC 439
\citep{Tisserand:2009kb}
and the other is MSX LMC 1795 \citep{Soszynski:2009ua}.
 \citet{Clayton:2011jw} studied another LMC R\,CrB star, HV 2671.
In our estimate, we integrate R\,CrB stars into carbon-rich AGB stars,
as they can not be distinguished from IR data only (Fig.\ref{Fig-38-8}).

AGB stars have one more chemical type, on top of oxygen-rich (M-type) and carbon-rich (C-type): S-type stars.
They have a carbon-to-oxygen ratio almost equal to unity. Their dust composition is diverse: FeO, FeS, silicate 
\citep{Zijlstra:2004dga, Smolders:2012to}.
The number of S-type stars is small ($\sim$1\,\% in the Galaxy) among AGB stars, so their contributions to the total
gas and dust budget is negligible.

\section{Discussions}
\subsection{Global gas and dust budget of the SMC}

In the SMC, the total gas injection rate from AGB stars, RSGs and SNe  is 
0.02--0.04\,\Msun\,yr$^{-1}$. This largely relies on the gas ejected
from SNe.

We compare this gas injected from SNe and evolved stars into the ISM
with gas consumed by star formation.
\citet{Kennicutt:1986bi} estimated the current SMC starformation rate of 
0.08\,\Msun\,yr$^{-1}$.
In the SMC ISM, 
gas consumed by the starformation exceeds
the gas injected from evolved stars and SNe (0.02--0.04\,\Msun\,yr$^{-1}$) into the ISM.
This is similar to what  has been  found in the LMC, though
the deficit in the LMC is much larger (about 0.1--0.2\,\Msun\,yr$^{-1}$).

The majority of the ISM gas in the SMC is present in the form of H{\small I}
and its mass is $4\times10^8$\,\Msun \citep{Stanimirovic:1999kx, Bolatto:2011et}.
A large ISM reservoir could sustain the star formation, despite a deficit in the gas injection rate, 
at least on a few Gyr time scale.
Unless there is a gas infall from an external source, the starformation rate will decline.

Similarly, SNe and AGB stars are important dust sources in the SMC.
SNe could contribute to a dust injection of 
2--110$\times10^{-5}$\,\Msun\,yr$^{-1}$  into the ISM,
while AGB stars and RSGs contribute about 1$\times10^{-5}$\,\Msun\,yr$^{-1}$ of dust mass.

 In the Milky Way, the lifetime of dust is estimated to be $6\times10^8$\,years
and $4\times10^8$\,years for carbonaceous and silicate dust grains, respectively \citep{Jones:1996bi}.
The lifetime of dust grains is constrained by their destruction by fast SN shocks
and the dust lifetime correlates with the SN Ia+II rate per surface area \citep{Dwek:1998js}.
The Galactic SN rate is estimated to be one in every 50 years \citep{Diehl:2006fr}.
This rate includes type Ib/c and II, but in practice type II dominates.
The type Ia rate in the Milky Way is about 0.3 per century \citep{Matteucci:2009ij},
so that the total SN Ia and II rate is about one event in every 45 years in the Milky Way.
The SMC has a rate of one event every 350 years, and in the LMC the rate is one event every 100 years 
\citep{Filipovic:1998vm}.
To calculate the SN rate per surface area, we use the half-light radius ($r_{1/2}$) of a galaxy as a measure of the surface area,
which we take from \citet{Tolstoy:2009cl}.
The estimated dust lifetime would be $5\times10^8$\,years
and $3\times10^8$\,years for carbonaceous and silicate dust grains, respectively in the LMC,
and $17\times10^8$\,years and $11\times10^8$\,yeas in the SMC.
In the SMC, the overall lifetime of the dust would be about 1.4\,Gyrs.
This assumes an almost constant SN rate and starformation rate over such a long period, 
which will be discussed in Sect.\,\ref{LMC-SMC}.
In the SMC, about $4\times10^4$--$1\times10^6$\,\Msun\, of dust has been accumulated over the 
1.4\,Gyr history of the SMC.

The estimated dust mass in the SMC ISM is $3\times10^5$\,\Msun\,
\citep{Leroy:2007da}.
If the dust lifetime in the SMC ISM is as long as 1.4\,Gyrs on average,
dust present in the SMC ISM may mainly originate from stellar dust.

The obvious uncertainty in this estimate is the lifetime of dust.
The half-light radius is an indicator of the galaxy size, but the LMC and the SMC are
irregular galaxies, and the actual SN rate per volume would not be so simple to estimate.
Furthermore, SNe might affect dust destruction in the SN vicinity only,
and dust in the remaining regions might survive. The uneven distribution of dust could bring diverse life-time of dust.
It seems that the dust lifetime largely depends on
the local condition of the galaxies, and more sophisticated calculations are needed.

It has been proposed that dust condensation in molecular clouds could be  a dominant source of ISM dust
in the Milky Way or high-redshift galaxies \citep{Tielens:2006vv, Draine:2009ur, Mattsson:2011if}.
These galaxies probably contain many molecular clouds within.
The SMC is known to have a low molecular content
\citep{Israel:1997tm}, 
so that dust condensation in the molecular clouds may not be as high as in the Milky Way.

The key to understand the evolution of ISM dust is to understand the formation and destruction
processes imposed by SNe.
It is now pausible to determine dust formation and destruction in SNe,
using the current missions/projects, such as Herschel, SOFIA and ALMA, but also future space missions \citep[JWST and SPICA; ][]{Tanaka:2012el}.
Also we need to find constraints for grain growth in molecular clouds based on observations.

\subsection{Comparison between the LMC and the SMC} \label{LMC-SMC}

In the SMC SNe are a more important gas source to the ISM than AGB stars and RSGs. 
The difference between them is a factor of 14--29.
In the LMC SNe have higher gas feedback rates than AGB stars and RSGs but the difference
is only by a factor of 4--9. AGB mass-loss rates have uncertainties of a factor of three at least 
\citep{Groenewegen:2007bi},
and the gas-to-dust mass ratio could have a factor of three uncertainties \citep{vanLoon:2000wa}.
The LMC difference is negligible, but the one in the SMC is not.
There is a difference in the main gas sources in the LMC and SMC.
The SMC has  more gas feedback from high-mass stars than the LMC.

The initial mass function \citep{Kroupa:2001vq} shows that equivalent mass should
be distributed to high-mass stars ($>$8\,\Msun) and low- and intermediate-mass stars (1--8\,\Msun).
At the event of stellar death, high-mass stars could return most of their mass into the ISM.
Many  low- and intermediate-mass stars end their lives as white dwarfs, and their masses are about 
0.6 \,\Msun \citep{1990ARA&A..28..103W}, thus 40--92\,\% of mass is returned to the ISM.
During a constant starformation rate period, the ratio of ISM gas feedback from high-mass stars
against low- and intermediate-mass stars should be about 3:2.
The measured ratio shows that SMC has an excess of gas feedback from high-mass stars
beyond the IMF.

The LMC and SMC have experienced more or less similar starformation histories
\citep{Harris:2004tg, Harris:2009bz}, but the SMC has an enhanced starformation history
by a factor of four in recent times in the last 12\,Myrs. That could result in a more efficient gas injection rate from SNe.

Additionally, the lower metallicity can cause lower dust driven winds, at least for oxygen-rich AGB stars
and red supergiants \citep{Bowen:1991kd, Marshall:2004ec}.
The metallicities of the LMC 
and the SMC are about half and the quarter of Solar \citep{Monk:1988vc}.
That could be another reason that there is a more efficient SN gas feedback in the SMC,
because of the relatively smaller contribution of oxygen-rich AGB stars and red supergiants.

\subsection{PAHs in the SMC}

 Carbon-rich AGB stars are considered to be the source of PAHs found the ISM 
 \citep{1989ApJS...71..733A, 1992ApJ...401..269C}.
PAHs are formed using C$_2$H$_2$ as a parent molecule in chemical reactions.
In the circumstellar envelope of carbon-rich AGB stars, carbon atoms are tied up in CO first,
and use up all oxygen.
and the excess carbon formed carbonaceous dust and carbon-bearing molecules.

 From our measured carbon-rich AGB gas injection rate, we can estimate the upper limit of PAH
mass injected from carbon stars.
The gas injection rate from carbon-rich AGB stars is  0.7$\times10^{-2}$\,\Msun\,yr$^{-1}$.
The C$_2$H$_2$ fractional abundance is approximately $10^{-5}$ in the Milky Way 
\citep{1996MNRAS.282L..21H}.
This can be has a factor of few higher,
depending on how the metallicities of the galaxies affect the amount of excess carbon in the stars
\citep{Matsuura:2005ej}, and to be predicted to be about $10^{-4}$\citep{Woods:2012hh}
in the Magellanic Clouds.
The maximum PAH injection rate, if all C$_2$H$_2$ is eventually converted into PAHs,
is 0.7$\times10^{-6}$\,\Msun\,yr$^{-1}$.

 The average fractional abundance of PAHs in the SMC is 0.6\,\% with respect to the total dust mass
 \citep{Sandstrom:2010ks}. The estimated PAH mass in the SMC to be
 about 1800\,\Msun.

 The lifetime of PAHs is shorter than amorphous carbon.
In the Milky Way, the estimated lifetime of PAHs is 1.4--1.6$\times 10^8$\,years \citep{Micelotta:2010dl}.
In the SMC, the SN rate is lower, and the lifetime of PAHs can be as long as a few $10^8$\,years.
The PAH injection rate from carbon-rich AGB stars is so low that
the expected PAH mass from AGB stars would be much lower than 100\,\Msun.
PAHs in the SMC require efficient in the ISM or in-falling into the SMC.
PAH formation needs ISM processing.
%or PAH mass fraction in the ISM may be overestimated,due to uncertainties in PAH emissivities.

\section{Conclusions}

We have measured the total gas and dust injection rates from AGB stars, red supergiants,
and also estimated these rates from supernovae.
The total gas injection from stellar deaths is about  2--4$\times10^{-2}$\,\mlu\, into the ISM.
This is slightly smaller than the current gas consumption in the ISM by starformation, which is $\sim$8$\times10^{-2}$\,\mlu.
The galaxy has a large gas reservoir the moment, so that it can sustain  high starformation rate at present.
Eventually the starformation rate is going to decline, unless an external source provides gas infall into the SMC.

AGB stars and red supergiants are  important sources of dust in the SMC.
Dust production in SNe is largely uncertain, so is the lifetime of  dust.
Within the current uncertainties in quantities, dust present in the ISM can be explained as being stellar in origin.

PAHs in the ISM can not be explained to have originated only from carbon-rich AGB stars.
The lifetime of PAHs is too short, compared with the supply from carbon-rich AGB stars.
PAHs require formation process in the ISM.

%\bibliographystyle{mn2e}
%\bibliography{references}{}
\bibliography{smc_ext}
%\bibliography{references}

\bibliographystyle{mn2e}

\label{lastpage}

\end{document}